\documentstyle[prl,eqsecnum,aps,floats,bibstyle]{revtex}

\begin{document}

\hyphenation{HERWIG}
\hyphenation{JETRAD}
\hyphenation{BFKL}

\preprint{FERMILAB-PUB-96/038-E}

\title{The Azimuthal Decorrelation of Jets Widely Separated in Rapidity}

% LIST_OF_AUTHORS.TEX                 02/22/96           
%
\author{                                                                        
%% names begin here                                                             
S.~Abachi,$^{14}$                                                               
B.~Abbott,$^{28}$                                                               
M.~Abolins,$^{25}$                                                              
B.S.~Acharya,$^{44}$                                                            
I.~Adam,$^{12}$                                                                 
D.L.~Adams,$^{37}$                                                              
M.~Adams,$^{17}$                                                                
S.~Ahn,$^{14}$                                                                  
H.~Aihara,$^{22}$                                                               
J.~Alitti,$^{40}$
% ...
\\                                                                              
\vskip 0.50cm                                                                   
\centerline{(D\O\ Collaboration)}                                               
\vskip 0.50cm                                                                   
}                                                                               
\address{                                                                       
\centerline{$^{1}$Universidad de los Andes, Bogot\'{a}, Colombia}               
\centerline{$^{2}$University of Arizona, Tucson, Arizona 85721}                 
\centerline{$^{3}$Boston University, Boston, Massachusetts 02215}               
\centerline{$^{4}$Brookhaven National Laboratory, Upton, New York 11973}        
\centerline{$^{5}$Brown University, Providence, Rhode Island 02912}             
\centerline{$^{6}$Universidad de Buenos Aires, Buenos Aires, Argentina}         
\centerline{$^{7}$University of California, Davis, California 95616}            
\centerline{$^{8}$University of California, Irvine, California 92717}           
\centerline{$^{9}$University of California, Riverside, California 92521}        
\centerline{$^{10}$LAFEX, Centro Brasileiro de Pesquisas F{\'\i}sicas,          
                  Rio de Janeiro, Brazil}                                       
\centerline{$^{11}$CINVESTAV, Mexico City, Mexico}                              
\centerline{$^{12}$Columbia University, New York, New York 10027}               
\centerline{$^{13}$Delhi University, Delhi, India 110007}                       
\centerline{$^{14}$Fermi National Accelerator Laboratory, Batavia,              
                   Illinois 60510}                                              
\centerline{$^{15}$Florida State University, Tallahassee, Florida 32306}        
\centerline{$^{16}$University of Hawaii, Honolulu, Hawaii 96822}                
\centerline{$^{17}$University of Illinois at Chicago, Chicago, Illinois 60607}  
\centerline{$^{18}$Indiana University, Bloomington, Indiana 47405}              
\centerline{$^{19}$Iowa State University, Ames, Iowa 50011}                     
\centerline{$^{20}$Korea University, Seoul, Korea}                              
\centerline{$^{21}$Kyungsung University, Pusan, Korea}                          
\centerline{$^{22}$Lawrence Berkeley National Laboratory and University of      
                   California, Berkeley, California 94720}                      
\centerline{$^{23}$University of Maryland, College Park, Maryland 20742}        
\centerline{$^{24}$University of Michigan, Ann Arbor, Michigan 48109}           
\centerline{$^{25}$Michigan State University, East Lansing, Michigan 48824}     
\centerline{$^{26}$Moscow State University, Moscow, Russia}                     
\centerline{$^{27}$University of Nebraska, Lincoln, Nebraska 68588}             
\centerline{$^{28}$New York University, New York, New York 10003}               
\centerline{$^{29}$Northeastern University, Boston, Massachusetts 02115}        
\centerline{$^{30}$Northern Illinois University, DeKalb, Illinois 60115}        
\centerline{$^{31}$Northwestern University, Evanston, Illinois 60208}           
\centerline{$^{32}$University of Notre Dame, Notre Dame, Indiana 46556}         
\centerline{$^{33}$University of Oklahoma, Norman, Oklahoma 73019}              
\centerline{$^{34}$University of Panjab, Chandigarh 16-00-14, India}            
\centerline{$^{35}$Institute for High Energy Physics, 142-284 Protvino, Russia} 
\centerline{$^{36}$Purdue University, West Lafayette, Indiana 47907}            
\centerline{$^{37}$Rice University, Houston, Texas 77251}                       
\centerline{$^{38}$Universidade Estadual do Rio de Janeiro, Brazil}             
\centerline{$^{39}$University of Rochester, Rochester, New York 14627}          
\centerline{$^{40}$CEA, DAPNIA/Service de Physique des Particules, CE-SACLAY,   
                   France}                                                      
\centerline{$^{41}$Seoul National University, Seoul, Korea}                     
\centerline{$^{42}$State University of New York, Stony Brook, New York 11794}   
\centerline{$^{43}$SSC Laboratory, Dallas, Texas 75237}                         
\centerline{$^{44}$Tata Institute of Fundamental Research,                      
                   Colaba, Bombay 400005, India}                                
\centerline{$^{45}$University of Texas, Arlington, Texas 76019}                 
\centerline{$^{46}$Texas A\&M University, College Station, Texas 77843}         
}                                                                               
%end                                                                            
\date{\today}

\maketitle

{\vspace{-0.2cm}
\begin{abstract}
This study reports the first measurement of the azimuthal decorrelation between 
jets with pseudorapidity separation up to five units.  The data were
accumulated using the D\O\ detector during the 1992--1993 collider run of the
Fermilab Tevatron at $\sqrt{s}=$ 1.8 TeV.  These results are compared to
next--to--leading order (NLO) QCD predictions and to two leading--log
approximations (LLA) where the leading--log terms are resummed to all orders in
$\alpha_{\scriptscriptstyle S}$.  The final state jets as predicted by NLO QCD
show less azimuthal decorrelation than the data.  The parton showering LLA Monte
Carlo {\small HERWIG} describes the data well; an analytical LLA prediction
based on BFKL resummation shows more decorrelation than the data.

\end{abstract}

\twocolumn
\input{psfig}

Correlations between kinematic variables in multijet events provide a simple way
to study the complex topologies that occur when more than two jets are present
in the final state~\cite{Stirling,DDS,high}.  For example, in dijet events the
two jets exhibit a high degree of correlation, being balanced in transverse
energy ($E_T$) and back--to--back in azimuth ($\phi$).  Deviations from this
configuration signal the presence of additional radiation.  Theoretically this
radiation is described by higher order corrections to the leading order graphs.
Using the four momentum transfer $Q^{2}$ in the hard scattering as the
characteristic scale and {\small DGLAP}~\cite{dglap} evolution in $Q^{2}$,
these corrections have been calculated analytically to NLO in perturbative
QCD~\cite{NLO,jetrad}.  In addition, they are approximated to all orders by
using a parton shower approach, like {\small HERWIG}~\cite{herwig} for example.
However, there can be more than one characteristic scale in the process.
Similar to deep inelastic lepton--hadron scattering at small Bjorken $x$ and
large $Q^{2}$, hadron--hadron scattering at large partonic center of mass
energies ($\hat{s}$) may require a different theoretical treatment.  Instead of
just resumming the standard terms involving $\ln Q^{2}$, large terms of the
type $\ln (\hat{s}/Q^{2})$ have to be resummed as well using the BFKL
technique~\cite{bfkl}.  Del Duca and Schmidt have done this and predict a
different pattern of radiation, which results in an additional decorrelation in
the azimuthal angle between two jets, as their distance in pseudorapidity
($\Delta\eta \sim \ln(\hat{s}/Q^{2})$) is increased~\cite{DDS}.

In this study, the jets of interest are those most widely separated in
pseudorapidity ($\eta = -\ln[\tan(\theta/2)]$, where $\theta$ is the polar
angle with respect to the proton beam).  The D\O\ detector~\cite{detector} is
particularly suited for this measurement owing to its uniform calorimetric
coverage to $|\eta| 
\mathrel{\rlap{\raise 0.4ex \hbox{$<$}}{\lower 0.72ex \hbox{$\sim$}}} 4.0$.
The uranium--liquid argon sampling calorimeter 
facilitates jet identification with its fine transverse segmentation ($0.1
\times 0.1$ in $\Delta\eta \times \Delta\phi$).  Single particle energy
resolutions are $15\%/\sqrt{E}$ and $50\%/\sqrt{E}$ ($E$ in GeV) for electrons 
and pions, respectively, providing good jet energy resolution.

The data for this study, representing an integrated luminosity of 83 nb$^{-1}$,
were collected during the 1992--1993 $\overline{p}p$ collider run at the
Tevatron with a center of mass energy of $\sqrt{s}=$ 1.8 TeV.  The hardware
trigger required a single  pseudo-projective calorimeter tower ($0.2\times0.2$
in $\Delta\eta \times \Delta\phi$) to have more than 7 GeV of transverse
energy.  This trigger was instrumented for $|\eta| < 3.2$.  Events satisfying
this condition were analyzed by an on-line processor farm where a fast version
of the jet finding algorithm searched for jets with $E_T > 30$ GeV.

Jet reconstruction was performed using an iterative fixed cone algorithm.
First, the list of calorimeter towers with $E_T > 1$ GeV (seed towers) was
sorted in descending order.  Starting with the highest $E_T$ seed tower, a
precluster was formed from all calorimeter towers with ${\cal R} < 0.3$, where
${\cal R} = \sqrt{\Delta\eta^{2}+\Delta\phi^{2}}$ was the distance between 
tower centers.  If a seed tower was included in a precluster, it was removed
from the list.  This joining was repeated until all seed towers become elements
of a precluster.  After calculating the $E_T$ weighted center of the precluster,
the radius of inclusion was increased to 0.7 about this center with all towers
in this cone becoming part of the jet.  A new jet center was calculated using
the $E_T$ weighted tower centers.  This process was repeated until the jet axis
moved less than 0.001 in $\eta$--$\phi$ space between iterations.  The final
jet $E_T$ was defined as the scalar sum of the $E_T$ of the towers; its
direction was defined using the D\O\ jet algorithm~\cite{angle}, which differs
from the Snowmass algorithm~\cite{Snowmass}.  If any two jets shared more than
half of the $E_T$ of the smaller $E_T$ jet, the jets were merged and the jet
center recalculated. Otherwise, any ambiguities in the overlap region were
resolved by assigning the energy of a given cell in the shared region to the
nearest jet.  Jet reconstruction was over 95\% efficient for jets with $E_T >
20$ GeV.  Jet energy resolution was 10\% at 50 GeV and jet position resolution
was less than 0.03 in both $\eta$ and $\phi$.

Accelerator and instrumental backgrounds were removed by cuts on the jet shape. 
The efficiency for these cuts was greater than 95\%.  Based on Monte Carlo
simulations, residual contamination from backgrounds was estimated to be less
than 2\%.  The  jet transverse energy was corrected for energy scale,
out--of--cone showering, and underlying event.  This correction was based 
on minimizing the missing transverse energy in direct photon
events~\cite{harry}.  Small pseudorapidity biases ($\delta\eta \leq 0.03$),
caused by the jet algorithm, were also corrected~\cite{daniel}.

\begin{figure}
\vspace{-8mm}
\centerline{\psfig{figure=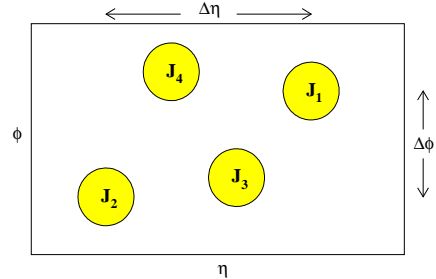,width=80mm}}
\vskip -1.4in
\caption{Typical event topology in multijet events.}
\label{ev config}
\end{figure}

A representative multijet event configuration is shown in Fig.~\ref{ev config}.
From the sample of jets with $E_T > 20$ GeV and $|\eta| \leq 3.0$, the two
jets at the extremes of pseudorapidity were selected ($J_{1}$ and $J_{2}$ in
Fig.~\ref{ev config}) for this analysis. One of these two jets was required to
be above 50 GeV in $E_T$ to remove any trigger inefficiency.  The 
pseudorapidity difference ($\Delta\eta = |\eta_{1} - \eta_{2}|$) distribution 
for events that pass the cuts is shown in Fig.~\ref{deta fig}.  In Fig.
~\ref{dphi fig}, the azimuthal angular separation, $1 - \Delta\phi/\pi$
($\Delta\phi = \phi_{1} - \phi_{2}$) is plotted for unit bins of $\Delta\eta$
centered at $\Delta\eta$ = 1, 3, and 5.  Since each distribution is normalized
to unity, the decorrelation between the two most widely separated jets can be
seen in either the relative decline near the peak or the relative increase in
width as $\Delta\eta$ increases.

\begin{figure}
\vspace{-5mm}
\centerline{\psfig{figure=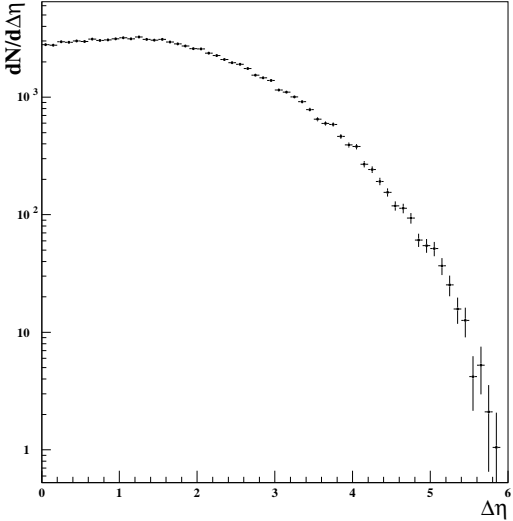,width=80mm}}
\caption{The pseudorapidity interval, $\Delta\eta = |\eta_{1} - \eta_{2}|$,
of the two jets at the extremes of pseudorapidity.  The coverage
extends to $\Delta\eta \sim 6$. The errors are statistical only.}
\label{deta fig}
\end{figure}

The decorrelation in Fig.~\ref{dphi fig} can be quantified in terms of 
the average value of $\cos(\pi-\Delta\phi)$~\cite{Stirling}.  Figure~\ref{c1
fig} shows $\langle\cos(\pi-\Delta\phi)\rangle$ vs. $\Delta\eta$.  For the
data, the error bars represent the statistical and point--to--point
uncorrelated systematic errors added in quadrature.  In addition, the band
at the bottom of the plot represents the correlated uncertainties of the energy
scale and effects due to hadronization and calorimeter resolution.  Also
shown in Fig.~\ref{c1 fig} are the predictions from {\small HERWIG}, NLO QCD as
implemented in {\small JETRAD}~\cite{jetrad}, and the BFKL 
resummation~\cite{DDS,DDS2}.  The errors shown for the three QCD predictions are
statistical only.

The systematic errors, especially the energy scale uncertainty, dominate the
statistical errors for all $\Delta\eta$ except for $\Delta\eta=5$. The jet
energy scale uncertainty is estimated to be 5\%.  The resulting uncertainty in
$\langle\cos(\pi-\Delta\phi)\rangle$ varied from 0.002 at $\Delta\eta =0$ to
0.011 at $\Delta\eta=5$. Since the out--of--cone corrections depended on the
pseudorapidity of the jet and may not be well understood at large
pseudorapidities, the full size of the out--of--cone showering correction was
included in the energy scale error band.  This size of this error was less than
0.013.  Uncorrelated systematic uncertainties due to the $\eta$ bias correction
and angular resolution were included.  This error was less than 0.002.  The jet
selection cuts introduced a systematic uncertainty less than 0.007, which is
independent of $\phi$ and $\eta$.  The uncertainty due to jet position
reconstruction was estimated by analyzing a subset of the data, specifically
events with a large $\Delta\eta$, using both Snowmass and D\O\ jet finding
algorithms; the differences in $\langle\cos(\pi-\Delta\phi)\rangle$ was less
than 0.002.

Comparison of theory with data requires the connection of partons with jets.
Since no attempt has been made to correct the data back to the parton level, the
the size of the hadronization and calorimeter resolution effects
were included as an additional systematic error.  These effects were estimated
using {\small HERWIG} with a detector simulation based on 
{\small GEANT}~\cite{geant}. Jets before hadronization were compared with jets
after both hadronization and detector simulation.  In both cases a cone jet
algorithm with a radius of 0.7 was used.  Jets reconstructed using partons and
particles produced indistinguishable results for $\langle\cos(\pi-\Delta\phi)
\rangle$; the calorimeter smearing effects, although negligible for $\Delta\eta
\leq 3$, were $\sim 0.02$ at $\Delta\eta=4$ and $\sim 0.03$ at $\Delta\eta=5$.
The size of these effects were included in the correlated systematic error band.

\begin{figure}
\vspace{-1cm}
\centerline{\psfig{figure=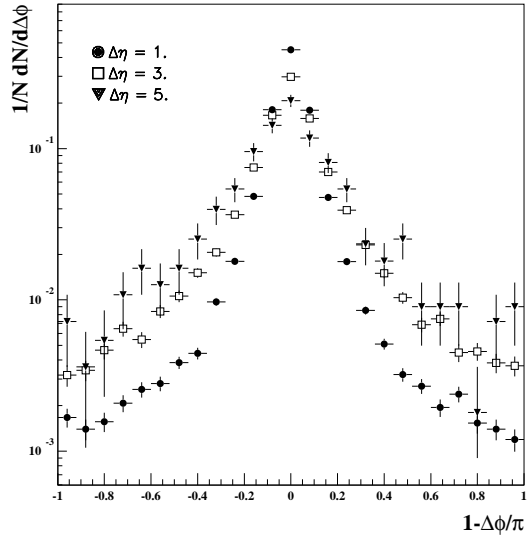,width=80mm}}
\caption{The azimuthal angle difference, $\Delta\phi = \phi_{1} - \phi_{2}$,
distribution of the two jets at the extremes of pseudorapidity plotted as
$1 - \Delta\phi/\pi$ for $\Delta\eta=1$, 3, and 5 ($0.5 < \Delta\eta < 1.5$,
$2.5 < \Delta\eta < 3.5$, and $4.5 < \Delta\eta < 5.5$).  The errors are
 statistical only.}
\label{dphi fig}
\end{figure}

Since NLO is the first order in perturbative QCD where decorrelation is
predicted, it may be sensitive to the choice of cutoff parameters (scales)
necessary in a perturbative calculation.  Similar effects have been seen in NLO
predictions of jet shape~\cite{jet shape} and topologies with jets beyond the
two body kinematic limit~\cite{walter}.  To estimate the size of these effects,
the renormalization and factorization scales in {\small JETRAD} were varied
simultaneously from $p_{T}^{max}/2$ to $2 p_{T}^{max}$, where $p_{T}^{max}$
is the transverse momentum of the leading parton.  The predictions for 
$\langle\cos(\pi-\Delta\phi)\rangle$ varied by less than 0.026.  The effect of
using different parton distribution functions (CTEQ2M~\cite{cteq},
MRSD$-$~\cite{mrs}, and GRV~\cite{grv}) produced variations in {\small 
JETRAD} that were less than 0.0025.  Since NLO QCD might be sensitive to the
jet definition, the jet algorithm angle definition study, previously done with
data, was repeated using {\small JETRAD}.  The difference between the Snowmass
and D\O\ definitions was smaller than 0.013 for all $\Delta\eta$.

The data in Fig.~\ref{c1 fig} show a nearly linear decrease in
$\langle\cos(\pi-\Delta\phi)\rangle$ with pseudorapidity interval.  For small
pseudorapidity intervals both {\small JETRAD} and {\small HERWIG} describe the
data reasonably well.  {\small JETRAD}, which is leading order in any
decorrelation effects, predicts too little decorrelation at large
pseudorapidity intervals.  The prediction of BFKL leading--log approximation,
which is valid for large $\alpha_{\scriptscriptstyle S}\Delta\eta$, is shown
for $\Delta\eta \geq 2$.  As the pseudorapidity interval increases, this
calculation predicts too much decorrelation.  Also shown in Fig.~\ref{c1 fig}
is the {\small HERWIG} prediction, where higher order effects are modeled with
a parton shower.  These predictions agree with the data over the entire
pseudorapidity interval range ($0 \leq \Delta\eta \leq 5$).

\begin{figure}
\vspace{-5mm}
\centerline{\psfig{figure=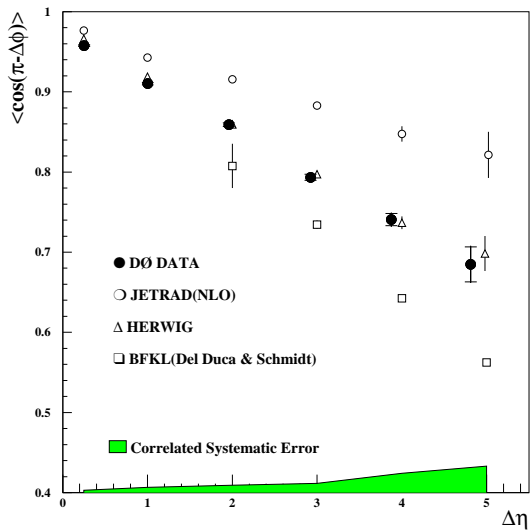,width=80mm}}
\caption{The correlation variable used in this analysis, the average value of
$\cos (\pi-\Delta\phi)$ vs. $\Delta\eta$, for the data, {\small JETRAD},
{\small
HERWIG}, and the BFKL calculations of Del Duca and Schmidt.}
\label{c1 fig}
\end{figure}

In summary, we have made the first measurement of azimuthal decorrelation as a
function of pseudorapidity separation in dijet systems.  These results have been
compared with various QCD predictions.  While the {\small JETRAD} predictions
showed too little and the BFKL resummation predictions showed too much
decorrelation, {\small HERWIG} describes the data well over the entire
$\Delta\eta$ range studied.

% Acknowledgement_paragraph.tex                             10/25/95
%
We appreciate the many fruitful discussions with Vittorio Del Duca and Carl
Schmidt.  
We thank the Fermilab Accelerator, Computing, and Research Divisions, and
the support staffs at the collaborating institutions for their contributions
to the success of this work.   We also acknowledge the support of
the U.S. Department of Energy,
the U.S. National Science Foundation,
the Commissariat \`a L'Energie Atomique in France,
the Ministry for Atomic Energy and the Ministry of Science and Technology 
   Policy in Russia,
CNPq in Brazil,
the Departments of Atomic Energy and Science and Education in India,
Colciencias in Colombia, 
CONACyT in Mexico,
the Ministry of Education, Research Foundation and KOSEF in Korea,
CONICET and UBACYT in Argentina,
and the A.P. Sloan Foundation.


\begin{references}
\bibitem{angle} The D\O\ jet algorithm defines the jet $E_T$ as 
                \[ E_T = \sum E_T^{\text{tower}},\]
                where the sum is over calorimeter towers in the jet cone.
                The angles are defined as
                \[ \theta_{\text{jet}} = \cos^{-1}\frac{E_{x}}
                {\sqrt{(E_{x})^{2}+(E_{y})^{2}+(E_{z})^{2}}} \] and 
                \[ \phi_{\text{jet}} = \tan^{-1} \frac{E_{y}}{E_{z}}, \]
                where
                \[ E_j = \sum E_i^{\text{tower}} {\text{for}}\ j = x,y,z.\]
\end{references}
\end{document}